\documentclass[10pt, conference]{IEEEtran}
\usepackage[left=0.625in,right=0.625in,top=0.75in,bottom=0.99in]{geometry}
\IEEEoverridecommandlockouts
\usepackage{cite}
\usepackage{amsmath,amssymb,amsfonts}
\usepackage{graphicx}
\usepackage{color}
\usepackage{subfig}
\usepackage{stackengine}
\usepackage{comment}
\usepackage{ulem}
\usepackage{tabularx}
\usepackage{bm}
\usepackage{dsfont}
\usepackage{mathtools}
\usepackage{lettrine}
\usepackage[font=scriptsize]{caption}

\usepackage[colorlinks, citecolor=blue]{hyperref} % to highlight references with colors

\usepackage[flushleft]{threeparttable} % http://ctan.org/pkg/threeparttable
\usepackage{booktabs}

\usepackage{textcomp}
\usepackage{xcolor}

\usepackage[linesnumbered, ruled]{algorithm2e}
\SetKwRepeat{Do}{do}{while}%

\SetCommentSty{mycommfont}
\usepackage{algpseudocode}

\def\BibTeX{{\rm B\kern-.05em{\sc i\kern-.025em b}\kern-.08em
    T\kern-.1667em\lower.7ex\hbox{E}\kern-.125emX}}
    
\usepackage{orcidlink}

\usepackage{xspace}
\usepackage{ifthen}
\usepackage[inline]{enumitem}

\usepackage{algcompatible}

\newboolean{showcomments}
\setboolean{showcomments}{true}
\ifthenelse{\boolean{showcomments}}
{
}

\usepackage{subfig}
\usepackage{fancyhdr}

\begin{document}

%\title{\huge Generative-Assisted DRL for SFC Provisioning via VNF Placement in 5G Core Network}
\title{\huge
GenAI Assistance for Deep Reinforcement Learning-based VNF Placement and SFC Provisioning in 5G Cores}

\author{
\IEEEauthorblockN{Murat Arda Onsu$^1$, Poonam Lohan$^1$, Burak Kantarci$^1$, Emil Janulewicz$^2$,}\\
\IEEEauthorblockA{\textit{$^1$University of Ottawa, Ottawa, ON, Canada}\\
\textit{$^2$Ciena, 383 Terry Fox Dr,
Kanata, ON K2K 2P5, Canada}\\
$^1$\{monsu022, ppoonam, burak.kantarci\}@uottawa.ca,~$^2$\{ejanulew\}@ciena.com}
}

%\author{\IEEEauthorblockN{ Pedro Enrique Iturria-Rivera\affmark[1], Marcel~Chenier\affmark[2] and Melike Erol-Kantarci\affmark[1], \IEEEmembership{Senior Member,~IEEE}}
%\IEEEauthorblockA{\affmark[1]\textit{School of Electrical Engineering and Computer Science, University of Ottawa, Ottawa, Canada}}  \affmark[2]\textit{NetExperience., Ottawa, Canada}\\
%Emails:\{pitur008, melike.erolkantarci\}@uottawa.ca,  marcel@netexperience.com  \vspace{-1em}}

%, Rick~Sommerville\affmark[2]
%rick.sommerville\} [2]

\maketitle
\begin{abstract}

Virtualization technology, Network Function Virtualization (NFV), gives flexibility to communication and 5G core network technologies for dynamic and efficient resource allocation while reducing the cost and dependability of the physical infrastructure. In the NFV context, Service Function Chain (SFC) refers to the ordered arrangement of various Virtual Network Functions (VNFs). To provide an automated SFC provisioning algorithm that satisfies high demands of SFC requests having ultra-reliable and low latency communication (URLLC) requirements, in the literature, Artificial Intelligence (AI) modules and Deep Reinforcement Learning (DRL) algorithms are investigated in detail. This research proposes a generative Variational Autoencoder (VAE) assisted advanced-DRL module for handling SFC requests in a dynamic environment where network configurations and request amounts can be changed. Using the hybrid approach, including generative VAE and DRL, the algorithm leverages several advantages, such as dimensionality reduction, better generalization on the VAE side, exploration, and trial-error learning from the DRL model. Results show that GenAI-assisted DRL surpasses the state-of-the-art model of DRL in SFC provisioning in terms of SFC acceptance ratio, E2E delay, and throughput maximization.

\end{abstract}


%\textbf{Index Terms} -- SFC Provisioning, NFV, VNF-Placement, Deep-Reinforcement Learning, Distributed Approach, Scalability, 5G and Beyond Network,  Priority Points

\section{Introduction} \label{sec:1}

Traditional network services and functions cause long product cycles and low service agility due to their strong dependability on specific hardware. Network Function Virtualization (NFV) brings novelty to communication technologies by decoupling software from physical devices and deploying functions on virtual machines residing on general-purpose hardware, such as data centers (DCs), to reduce OPEX and CAPEX while increasing flexibility \cite{4}. Service Function Chaining (SFC), in the NFV context, can be defined as a sequence of Virtual Network Functions (VNF) to deliver services such as Cloud Gaming (CG), Augmented Reality (AR), Voice over IP (VoIP), Video Streaming (VS), Massive IoT (MIoT), and Industrial 4.0 (Ind 4.0) \cite{2} \cite{12}. SFC provisioning requires satisfying all VNFs in its chain in proper order and establishing a packet transfer. However, SFC provisioning encounters several critical challenges, including efficient resource allocation, sequential execution of VNFs, high demands, and E2E delay limitations.

Machine Learning (ML) and  Deep Learning (DL) methods have been highly investigated in the literature to overcome these challenges. Deep Reinforcement Learning (DRL) has made a significant breakthrough and impacted various areas due to its advanced methodology. Combining both DL and Reinforcement Learning (RL), which provides exploration and trial and error methods by using DL model architecture; DRL has also been used in optimal VNF placement and SFC provisioning in literature \cite{3} \cite{5}. The goal of DRL is to maximize the expected reward by performing actions considering its current state. However, this methodology can struggle at the beginning of the training process or generalization of the environment, which makes it require additional assisted models such as a generative approach \cite{9}. Generative models, such as Variational Autoencoder (VAE) or Generative Adversarial Network (GAN), can assist DRL in numerous ways, such as additional unseen data generation, understanding action-value or state-value distributions, state representations, and so on \cite{1} \cite{8}.

In this research, we enhance the model from our prior study \cite{onsu2024unlocking}, which employed a DRL model with multiple input layers and priority points to unlock reconfigurability in SFC provisioning. The current work integrates a generative-assisted approach, deploying a VAE \cite{14}, to further refine the model's performance and adaptability. In the previous work, the data center (DC) is selected using the priority points, and selected DC information alongside overall system states is given to the DRL model. Then, the DRL model performs a VNF placement action on the selected DC. However, since the selection of the appropriate DC is solely based on a priority point, other critical aspects like available computational resources, network congestion, and task completion time are often overlooked. This may result in suboptimal DC selection, leading to inefficient resource allocation. Therefore, this study aims to use a VAE-based generative AI-assistance DRL model (GenAI-DRL) where the network and each DC information are given to the generative part, and it calculates the state-value function of each DC; then, the DC with maximum value will be selected for DRL to perform an action. Training of the VAE is performed by the dataset of DCs' initial state and next state after the DRL model's action inspired by \cite{13}.  The main contributions of this work are summarized as follows:

\begin{itemize}
    \item Improving the earlier DRL model with the generative approach with VAE where DCs in the network are selected by their value function created by the generative model.
    \item Advanced training of VAE by using its current state and next state after the DRL model performs an action. The dataset is collected during the simulation runtime from previous research's VNF provisioning model and randomly selecting the DC algorithm. After each DC is selected randomly, their next states and values are calculated and pushed forward alongside with current state to the dataset of VAE. 
    \item Detailed analyses are done for E2E latency, SFC acceptance ratio, and overall network throughput.  
\end{itemize} 

The rest of the paper is organized as follows. Section \ref{sec:2} presents related works. Section \ref{sec:3} describes the system model and problem formulation. The proposed GenAI-assisted DRL model is explained in Section \ref{sec:4}. Section \ref{sec:5} provides numerical results and discussions, and Section \ref{sec:6}  concludes the paper.

\section{Related Works} \label{sec:2}

Several studies focus on SFC provisioning and
VNF placement in the literature. In study \cite{6}, PSVShare is proposed for service placement problems in edge computing considering priority-based, least-cost, and
resource-efficient direction. On the other hand, researchers in \cite{7} adapt Q-learning to find the optimal SFC path considering the resource utilization of VNF placement in 5G. DRL algorithm based on Deep-Q-Network (DQN) is utilized in \cite{10} for maximizing QoE while meeting QoS requirements in NFV-enabled networks. The DRL-Based method is also used in \cite{11} for the VNF cooperative scheduling framework with priority-weighted delay to handle the issues of VNF queueing waiting and resource imbalances. Double Deep Q Network-based VNF Placement Algorithm is proposed in \cite{5} for optimal VNF placement, which minimizes the rejection rate and E2E latency while maximizing the throughput.

Furthermore, generative models and robust RL methods have been investigated in the literature for effective resource management. For example, researchers in \cite{1} introduce a  GAN-powered Deep Distributional Q-Network (GAN-DDQN) to mitigate randomness in resource allocation by learning the action-value distribution for each service slice in 5G networks. Here, the GAN approximates the distribution using a generative approach to model the randomness and noise embedded in the resource allocation problem, enabling the model to adapt to dynamic environments. Study \cite{8} combines curiosity-driven exploration with the VAE forming a DQN-CVAE model that generates more diverse and high-quality samples. This improves sample efficiency and enhances the training process of the Q-network. Another Approach, The Dreaming Variational Autoencoder (DVAE) presented in \cite{13}, generates action-value distributions by encoding state-action pairs into a latent space and decoding them into probable future states. This approach allows the model to predict future states based on past actions and store these predictions in a replay buffer for further training. 

While DL and DRL methodologies are explored in the literature for SFC provisioning, generative approaches offer the potential for more robust training and optimization, particularly in complex systems where exploring and training AI agents effectively is challenging. However, the integration of DRL with generative approaches remains underexplored in existing research. Therefore, this study will investigate GenAI-assisted DRL to enhance the performance of the AI agent.

\begin{table} 
\centering
\caption{Service Function Chain (SFC) characteristics in 5G Core Network\cite{2}}\fontsize{6.5}{7.7}\selectfont
\begin{tabular}{|p {1.4cm}|p {1.6cm}|p {1cm}|p {1cm}| p{1cm}|} 
 \hline
 \textbf{SFC Request}&\textbf{VNF Sequence} &\textbf{Bandwidth (Mbps)} &\textbf{E2E delay (msec)} &\textbf{Request Bundle} \\  
 \hline
  Cloud Gaming (CG)  & NAT-FW-VOC-WO-IDPS  & 4 & 80  & [40-55] \\
  \hline
  Augmented Reality (AR) & NAT-FW-TM-VOC-IDPS & 100 & 10  & [1-4] \\
  \hline
  VoIP & NAT-FW-TM-FW-NAT & 0.064 & 100 & [100-200] \\
  \hline
  Video Streaming (VS)& NAT-FW-TM-VOC-IDPS & 4  & 100  & [50-100] \\
  \hline
  MIoT & NAT-FW-IDPS  & [1-50] & 5 & [10-15] \\
  \hline
  Ind 4.0 & NAT-FW & 70 & 8 & [1-4] \\
 \hline
\end{tabular}
\label{tab:1}
\end{table}

\begin{figure*}
    \centering
    \includegraphics[width=0.75\linewidth]{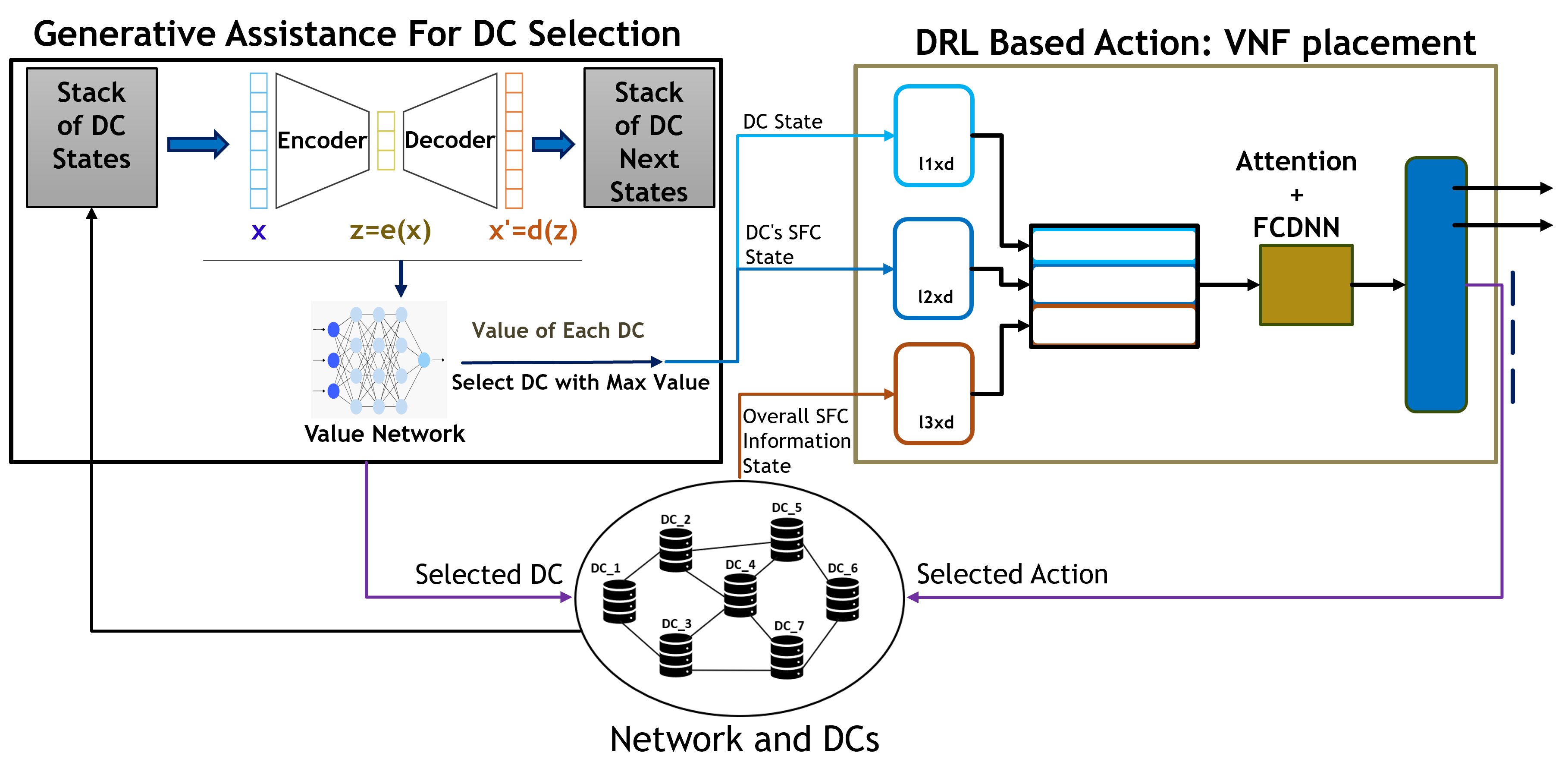}
    \caption{(\textbf{GenAI-DRL}) System model Illustration with various DCs and GenAI-Assisted DRL Agent}
    \label{fig: system model}
\end{figure*}

\section{System Model and Problem Formulation} \label{sec:3}

%\subsection{System Models and Environment Components}

\figurename \ref{fig: system model} illustrates the network environment, including VNF instance (VNFI)-enabled DCs, $\mathcal{D}$, logical links between DCs, $\mathcal{L}$, and the GenAI-assisted DRL agent for SFC provisioning. %AI module aims to satisfy SFC requests in the network via the proper sequential VNF placement as per their order in the chain to the proper DCs.
Each DC, $i\in \mathcal{D}$, has limited computational and storage capacity denoted by $\mathcal{C}_i$ and $\mathcal{S}_i$, respectively. The BW capacity of the logical link between DCs $i$ and $j \in \mathcal{D} $ is denoted by $B_{ij}$ Mbps. The set of supported SFC requests is denoted by $S=\{\text{CG, AR, VS, VoIP, MIoT, Ind 4.0}\}$. Each SFC requests $s \in S$ is composed of ordered VNF sequences $\mathcal{V}^s=(V_1^s\rightarrow V_2^s,\rightarrow ... V_{N_s}^s)$ and have attributes of bandwidth requirement, $B^s$ (Mbps), E2E delay, $D^s$ (msec), requests bundle size, $\Lambda^s$, as represented in Table \ref{tab:1}. To satisfy the SFC request, the AI model should place and process all VNFs to proper DCs as per its chain order before exceeding the E2E delay. The set of all VNFs is denoted by $\mathcal{V}=\{\text{NAT, FW, VOC, TM, WO, IDPS}\}$. The placement of each VNF, $v \in \mathcal{V}$, requires some storage, computational capacity, and processing time of DC denoted by $c^v$, $s^v$, and $t_p^v$, respectively. On each DC multiple same and different types of VNFs can be installed while following DC's storage and computational capacity constraints. Let $x_i^v$ denote the number of VNF $v\in V$ installed on DC $i \in \mathcal{D}$, then the storage and computational capacity constraints are represented by $C1: \sum_{v \in V}x_i^v s^v \leq \mathcal{S}_i, \; \forall i \in \mathcal{D}$ and  $C2: \sum_{v \in V}x_i^v c^v \leq \mathcal{C}_i, \; \forall i \in \mathcal{D}$, respectively. It is assumed that each VNF of an SFC request can be processed by only one DC having that VNF type installed and available on it for allocation till its processing time duration. Let $\Delta_i^{V_m^s}$ represent the binary variable which is set to $1$ if the $m^{th}$ VNF of SFC $s\in S$ is allocated at DC $i\in \mathcal{D}$, otherwise it is set to $0$. With this, the above assumption can be written as the following constraint $C3: \sum_{i \in \mathcal{D}} \Delta_i^{V_m^s} \mathds{1}(x_i^{V_m^s}>0)=1, \; \forall s \in S, \; \forall m=\{1,2,...,N_s\}$. The constraint related to logical link  BW capacity stating that the BW resources occupied by SFC requests should not exceed the capacity of logical links is represented as follows $C4: \sum_{\Lambda_s \forall s \in S} \sum_{m=1}^{N_s-1}\Delta_i^{V_m^s} \Delta_j^{V_{m+1}^s} B^s\leq B_{ij}, \; \forall i,j \in \mathcal{D}, \; i\neq j $. At last, the sum of the processing delay of all VNFs in the chain, $t_p^s$, and propagation delay from source to destination, $t_g^s$ should be less than the E2E delay tolerance limit of the SFC request i.e. $C5: t_P^s+t_g^s \leq D^s, \; \forall s\in S$.  Hence, while fulfilling the above-stated five constraints, the goal of this work is to maximize the acceptance ratio (AccRatio) of all types of SFC requests received in bundles with optimally selecting the DCs for VNF placement and allocation to SFC requests which can be represented mathematically as follows:
\begin{align*}
    &(\mathcal{P}):\; \underset{\mathbf{x},\boldsymbol{\Delta}}{\text{maximize}}\; R_a={\sum_{s\in S}R_s}/{\sum_{s\in S}\Lambda_s}\\
    &\text{s.t.:}\; C1,C2,C3,C4,C5.
\end{align*}
Where $R_a$ denotes the AccRatio defined by the ratio of the total number of all types of accepted requests, $R_s\; \forall s\in S$ to the total number of generated requests of all types of request bundles $\Lambda_s\; \forall s\in S$ .

%Further, it is considered that all types of SFC requests coming to the network are in the form of request bundles, i.e., multiple different types of users’ requests are considered to address massive demands the network should be able to handle within their E2E delay limit. Therefore, an efficient and optimum SFC provisioning model should be applied to the system to maximize the SFC request acceptance ratio and throughput while minimizing the E2E delay and rejection. 

To solve this combinatorial and nonconvex problem, we have proposed a novel GenAI-assisted DRL-based SFC provisioning solution. In the proposed solution, a generative VAE model is used to select the most important DC to perform an action. The selected DC's information, alongside overall network information and SFC requests, is given to the DRL-based VNF placement model, and the DRL agent decides what kind of action to apply on that DC. GenAI model helps the state representation and generalization of each DC and its embedded representation is pushed forwarded to the value network where each DC is assigned a proper importance value. DRL model selects the most important DC considering these values and performs the action on the selected DC. Every time the AI agent takes action, the system model updates its parameters and changes state information.

%\subsection{Problem Formulation}

\section{{GenAI}-Assisted DRL Based SFC Provisioning} \label{sec:4}

The proposed GenAI-assisted DRL agent for SFC provisioning is divided into two parts: generative VAE and Deep Q-network (DQN). In this algorithm, GenAI module selects the DC, and the selected DC's information is given to the DQN alongside environmental information for VNF placement. VAE is a GenAI model designed for unsupervised learning tasks, particularly in high-dimensional data contexts. It is a type of autoencoder, but with a probabilistic twist, and composed of two main components: the encoder and the decoder. The encoder maps input data $x$ into a latent space represented by a probability distribution, typically a Gaussian distribution. Instead of mapping directly to a fixed latent representation, it learns the parameters of this distribution, such as the mean $\mu$ and the variance $\sigma^2$, capturing uncertainty about the latent variables. The encoded latent representation $z$ is then sampled from this learned distribution. On the other hand, The decoder reconstructs the original input data from the sampled latent variable  $z$. It aims to generate data points that are similar to the original input, ensuring that the latent space captures meaningful variations in the data. In this research, DCs' current state in the simulation is given to the VAE, and the model, instead of generating the original input sequence, tries to estimate the DC's next state after the DRL module performs the VNF placement action on each of them, so it helps generalization and reaching the optimum DC.

\subsection{Generative AI (GenAI) Assistance Model}

\begin{figure}
    \centering
    \includegraphics[width=0.60\linewidth]{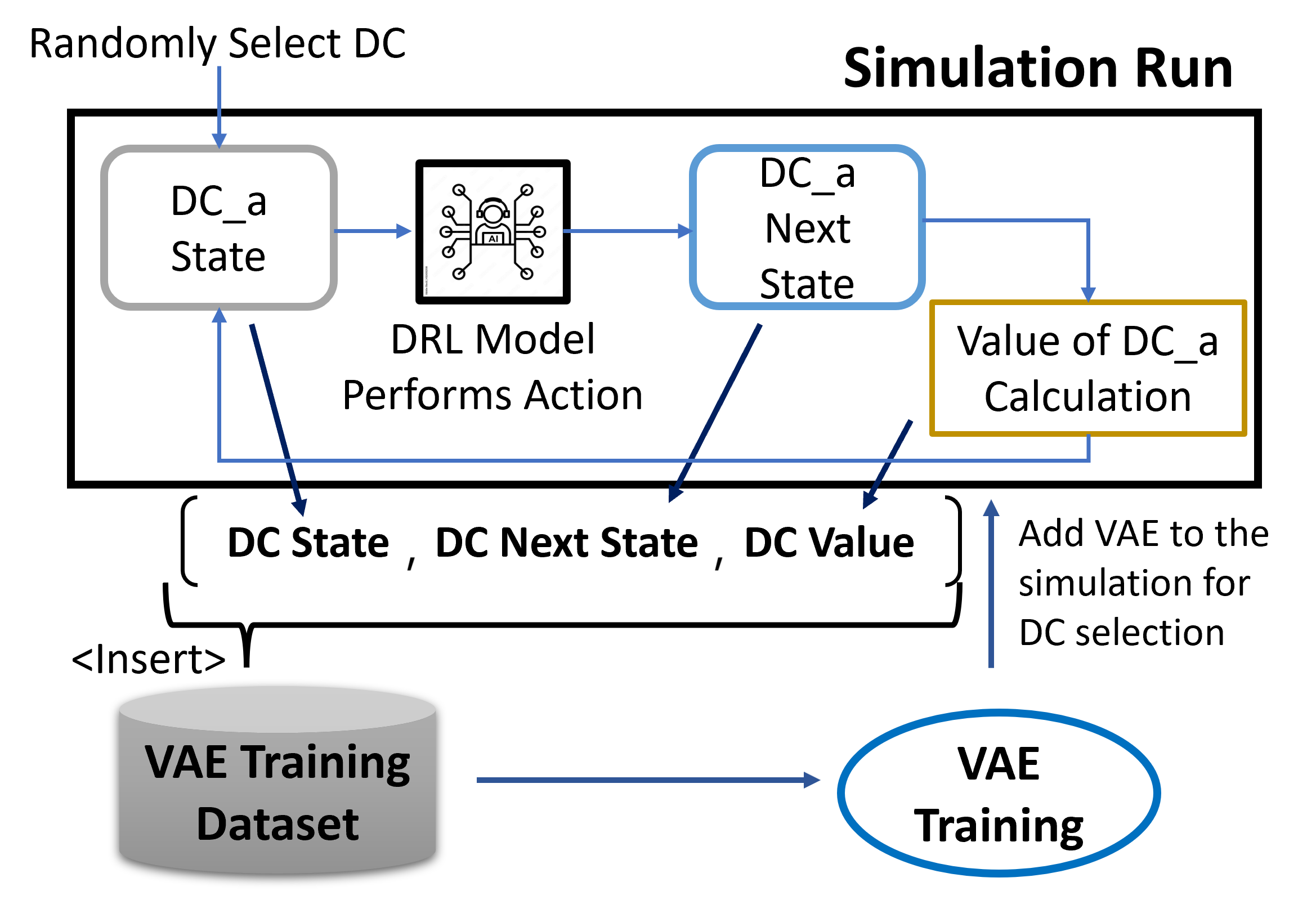}
    \caption{ VAE model training flow}
    \label{fig: VAE}
\end{figure}

The generative model is trained using two key objectives: minimizing the reconstruction loss, which measures how well the generated data matches the next state after DRL module performs an action, and the Kullback–Leibler (KL) divergence between the learned latent distribution and a prior distribution, typically a standard normal distribution. The state of the DCs includes its resource availability, installed VNF functions, and the important SFC features such as how many SFCs of which type have a source on that DC, which SFC has minimum E2E remaining time, BW availability, and whether is it possible to allocate the SFC's VNF  on that DC. This information is given to the VAE's encoder, and the decoder part tries to generate the next state if the DRL model performs VNF placement action. Model training is illustrated in  \figurename \ref{fig: VAE}.

The dataset for generative VAE is collected during the simulation runtime. In that runtime, DC selection is performed on a random basis to improve the diversity and exploration. Then, its information is stored as DC's current state for VAE training. After DC is selected, already trained DRL from previous work \cite{onsu2024unlocking} is utilized for selecting the proper actions (VNF allocation, VNF uninstallation, Idle wait), and the environment is updated. Based on these actions, the DC’s state is updated, and its value is calculated by taking into account the SFCs that the DRL module aims to satisfy, along with factors such as their source, destination, remaining E2E latency, and the availability of resources and bandwidth. The DC's next state and value, alongside the DC's current state, are added to the dataset, and once the dataset is large and diverse enough, the simulation stops, and the generative VAE model starts training. Training is performed in two phases: VAE training and Value Network Training. In VAE training, the model is fed by the DCs' current state, and it tries to generate the next state while representing them in the embedded space.Upon completion of VAE training, the embedded representations of the DCs, along with their values, are used to train the Value Network.

Once the VAE model and the Value Network have been trained, they are used with DRL for SFC provisioning in the simulation runtime. Overall, DCs' current states are given to the VAE's encoder part, where their embedded representations are created. Then, these embedded representations are forwarded to the Value Network, and each of them is assigned a value that shows DC's importance. The DRL model selects the DC with the maximum value and performs an action on the selected DC.

\subsection{Deep Reinforcement Learning (DRL) Module}

The trained DRL module that performs VNF placement action on the selected DC for SFC provisioning is used from our previous work \cite{onsu2024unlocking}. A Markov Decision Process (MDP) is utilized for modeling decision-making in situations which has the components of state, action, next state and reward. The main aim of MDP is to find the policy that maximizes the cumulative reward value, and the DQN algorithm is applied to address the MDP. In the advanced DRL technique, DQN, deep learning is used to estimate the optimal action-value function, commonly referred to as the Q-value. 

The reward is scalar feedback indicating the agent's performance for each action, where penalties are assigned for suboptimal decisions. An optimal action, like successfully fulfilling an SFC request by placing all its VNFs, results in a positive reward of $+2$. Conversely, dropping an SFC request incurs a penalty of $-1.5$, which is slightly less severe than the reward. Additional penalties are given for actions such as uninstalling essential VNFs $(-0.5)$ and selecting invalid action $(-1)$ that leaves the state unchanged, keeping the model in the same action until the next step. The actions in the DQN module include placing/allocating a VNF, uninstalling a VNF from the selected DC, and idle waiting, which is also considered a valid action.
 The number of neurons in the output layer of the DRL corresponds to the total number of possible actions and is defined as [$2*|\mathcal{V}|+1$], where $|\mathcal{V}|$ denotes the number of VNF types.

Finally, States represent the information the DRL agent knows about the environment at a given time, and these are given as inputs of the module. DQN module is constructed via a fully connected deep neural network (FCDNN) layers architecture with 3 input layers, of which the first input layer is related to the selected DC's resources and the VNFs installation situation on it, the second layer corresponds to SFC information on that DC, and the third input layer takes the SFC information for the overall network.  These layers include different numbers of input neurons, considering the states' features, but the same number of output neurons to generate the output with the same size. Then, the outputs of these layers are concatenated to form one instance forwarded to the attention layer to emphasize the important features. Then, it passes through several hidden layers until reaching the final layer. The number of neurons in the final layer is equal to the number of action types of the DQN, and the model performs an action considering the output value.

\subsection{SFC Provisioning Algorithm}

\begin{algorithm}    
\caption{SFC Provisioning via GenAI-DRL} \label{alg:vnf-placement}
\fontsize{7.5}{7.7}\selectfont
\KwData{Overall\_SFCs, Overall\_Network}
\KwResult{DC\_States, DC\_Next\_States, DC\_Values}
\SetKwFunction{GenerativeAssistanceModel}{GenAI\_Deploying}
\SetKwFunction{DQNModelDeploying}{DQN\_Model\_Deploying}
\SetKwFunction{MaxValueDC}{Max\_DC}
\SetKwFunction{GenerateStates}{Generate\_States}
\SetKwFunction{CalculateValue}{Calculate\_Value}
 $Gen\_Model \gets \GenerativeAssistanceModel()$\;
 $DQN\_SFC\_Provisioning\_Model \gets \DQNModelDeploying()$\;
 $DC\_States \gets []$, $DC\_Next\_States \gets []$, $DC\_Values \gets []$\;
\While{$Overall\_SFCs.SFCs\_exists$}
{
    \ForEach{$DC \in Overall\_Network.DCs$}
    {
         $DC\_States.append(DC.get\_state())$
    }
     $DC\_Repr \gets Gen\_Model.VAE.Encode(DC\_States)$\;
     $DC\_Values \gets Gen\_Model.Value\_Network(DC\_Repr)$\;
     $DC\_Selected \gets \MaxValueDC(DC\_Vals, Overall\_Network.DCs)$\;
     $DQN\_State \gets \GenerateStates(DC\_Selected,$    \\\phantom{for}\phantom{for}\phantom{for}\phantom{for}\phantom{for}\phantom{for}\phantom{for}\phantom{for}\phantom{for}\phantom{for}$Overall\_Network, Overall\_SFCs)$\\
     $DQN\_SFC\_Provisioning\_Model.Action(States\_for\_DQN)$\\
    $Overall\_Network.update()$\;
    \ForEach{$(id, DC) \in Overall\_Network..DCs\_with\_id$}
    {
      $DC\_Next\_States.append(DC.get\_states())$\;
      $DC\_Values.append(\CalculateValue(DC, DC\_States[id]))$
    }
}
\end{algorithm}

In Algorithm \ref{alg:vnf-placement}, overall SFC information and network components such as DC's logical links and resources are given as a parameter, and the algorithm runs until there are no SFC requests in the system. At the beginning of the process, the GenAI model is deployed after being trained in the previous run, and the deployment of the DQN model follows it. In step 4, the simulation starts handling the SFC request, and in the first phase, all DCs' current states are stored in $DC\_State$ (Steps 5-7). These states' information is first sent to the encoder parts of the VAE in the GenAI model (step 8), and their embedded representations are collected. Then, these representations are forwarded to the value network (step 9), and each DC is assigned a specific value. Afterward, DQN selects the DC with maximum value (step 10) and performs a VNF placement action on it by taking the SFC and source information of the selected DC and overall SFC information (steps 11-13). After the VNF placement is completed, system components such as DC start updating, and new DC state information alongside their calculated values are stored in  $DC\_Next\_States$ and $DC\_Values$ are stored for another training and testing for generative modeling (Steps 14-18). The algorithm is concluded after handling all SFCs in the system.

\section{Numerical Results} \label{sec:5}

\subsection{Simulation Setup and Training}

The simulation model for the SFC provisioning algorithm is developed in Python. The network consists of DCs, logical links, and SFC requests that include specific VNF chains. Each DC is equipped with 2 TB of storage, CPUs ranging from 12 to 120 GHz, and 256 GB of RAM, while the logical links offer 1 Gbps bandwidth. SFC requests are generated by clients as bundled requests (see Table \ref{tab:1}) and are sent to the network for processing \cite{onsu2024platform}. The simulation generates the six most common SFCs: CG, AR, VoIP, VS, MIoT, and Ind 4.0, as outlined in \cite{2} characterized by specific bandwidth requirements, E2E delay, request bundle ranges, and VNF sequences, as detailed in Table \ref{tab:1}. The VNF sequences include components such as Network Address Translation (NAT), Firewall (FW), Video Optimization Controller (VOC), Traffic Monitor (TM), WAN Optimizer (WO), and Intrusion Detection Prevention System (IDPS). To satisfy an SFC request, its VNFs must be processed in the proper order. The vCPU, RAM, storage, and processing times for each VNF are sourced from \cite{2}. Efficient VNF placement, installation, and allocation are crucial for meeting the E2E delay constraints of SFC requests and improving network performance, particularly in terms of AccRatio.

Pre-trained DQN is used for SFC provisioning via VNF placements from previous work \cite{onsu2024unlocking} where the model is trained with 2-4 DCs environment with 350 updates every 20 episodes. For GenAI model training, the simulation runs with pre-trained DQN for SFC provisioning, and DCs are selected randomly during the runtime. After each action of DQN, DC's next states, alongside the current state and calculated value, are added to the dataset, and generative model training is performed by using this dataset. SFC requests are generated randomly within their bundle sizes (see Table \ref{tab:1}). A request is successfully fulfilled if all VNFs in its chain are processed within the required time; otherwise, it is dropped due to resource constraints or failure to meet E2E delay requirements.

\begin{figure}
    \centering
    \includegraphics[width=1.\linewidth]{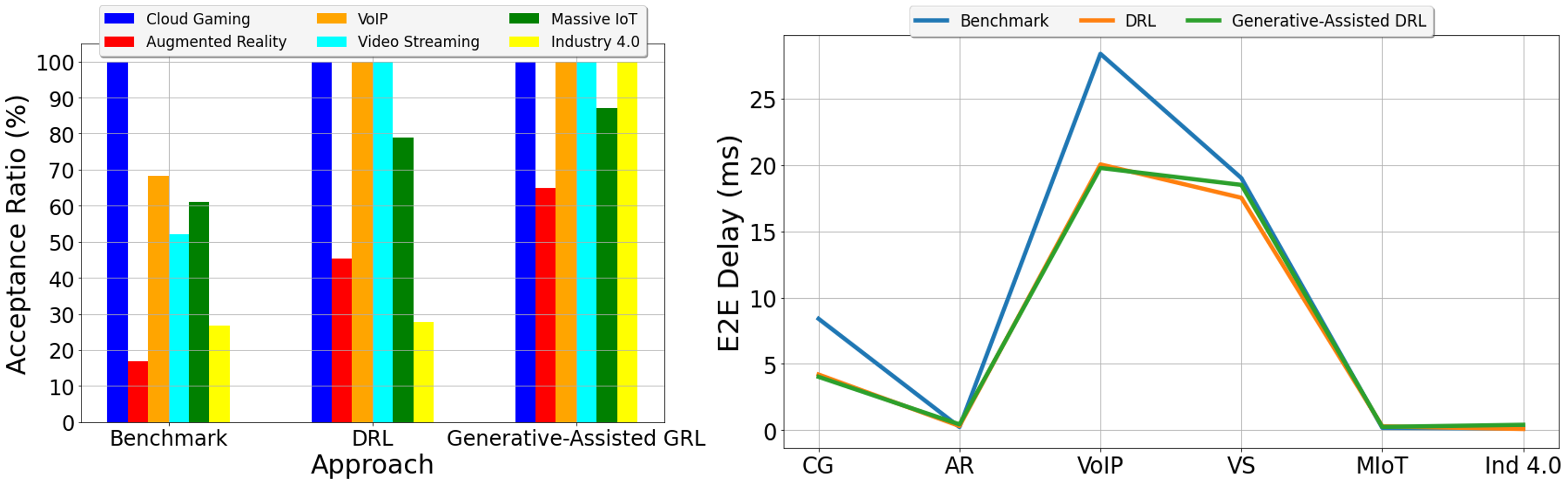}
    \caption{SFC AccRatio and E2E Delay Under Different Algorithms }
    \label{fig: res 1}
\end{figure}

\subsection{Testing and Results}

To measure the proposed approach's performance and superiority over the previous methods and benchmark algorithm, the following metrics have been utilized: E2E delay, throughput, overall SFC AccRatio, and AccRatio of each SFC type under different network configurations with different amounts of demands.  In \figurename \ref{fig: res 1}, the left diagram compares the performance of the proposed Gen-AI DRL algorithm with the baseline rule-based heuristic approach and DRL in terms of AccRatio with 6 DC environment. Although the comparison is percentage-based, each SFC type has a different bundle size, as outlined in Table \ref{tab:1}. According to the results, the heuristic-based approach achieves minimum AccRatio, while it is followed by DRL and the proposed GenAI-DRL algorithm. VS and VoIP requests are not satisfied completely only in a benchmark, with 53\% and 60\%, while the Ind 4.0 satisfaction rate is below 30\% for both DRL and heuristic approaches. AR reaches minimum AccRatio in heuristic approached with around 18\% due to strict E2E delay limits, while it is increased to 45\% and 64\% in DRL and GenAI-DRL algorithm. The proposed algorithm 100\% satisfies CG, VoIP, VS, and Ind 4.0 while MIoT AccRatio is 88\%. Moreover, CG has the highest AccRatio for all algorithms due to its 80 ms delay tolerance, which is lower than that of VS and VoIP (100 ms), making low E2E delay tolerant requests prioritized. On the other hand, the right part of the figures compares the E2E latency of this runtime, and AR, VS, MIoT, and Ind 4.0 have the minimum E2E latency for all algorithms since having minimum E2E delay limits makes them be processed immediately. The benchmark model reaches maximum latency. Though the DRL model and GenAI-DRL model's E2E latency are almost similar and sometimes superior to each other, it must be noted that E2E delay results are calculated considering the accepted SFC requests, and dropped SFCs are not used in this calculation. %Therefore, the proposed algorithm surpasses DRL for combined SFC AccRatio and E2E delay results. 

\begin{figure}
    \centering
    \includegraphics[width=.90\linewidth]{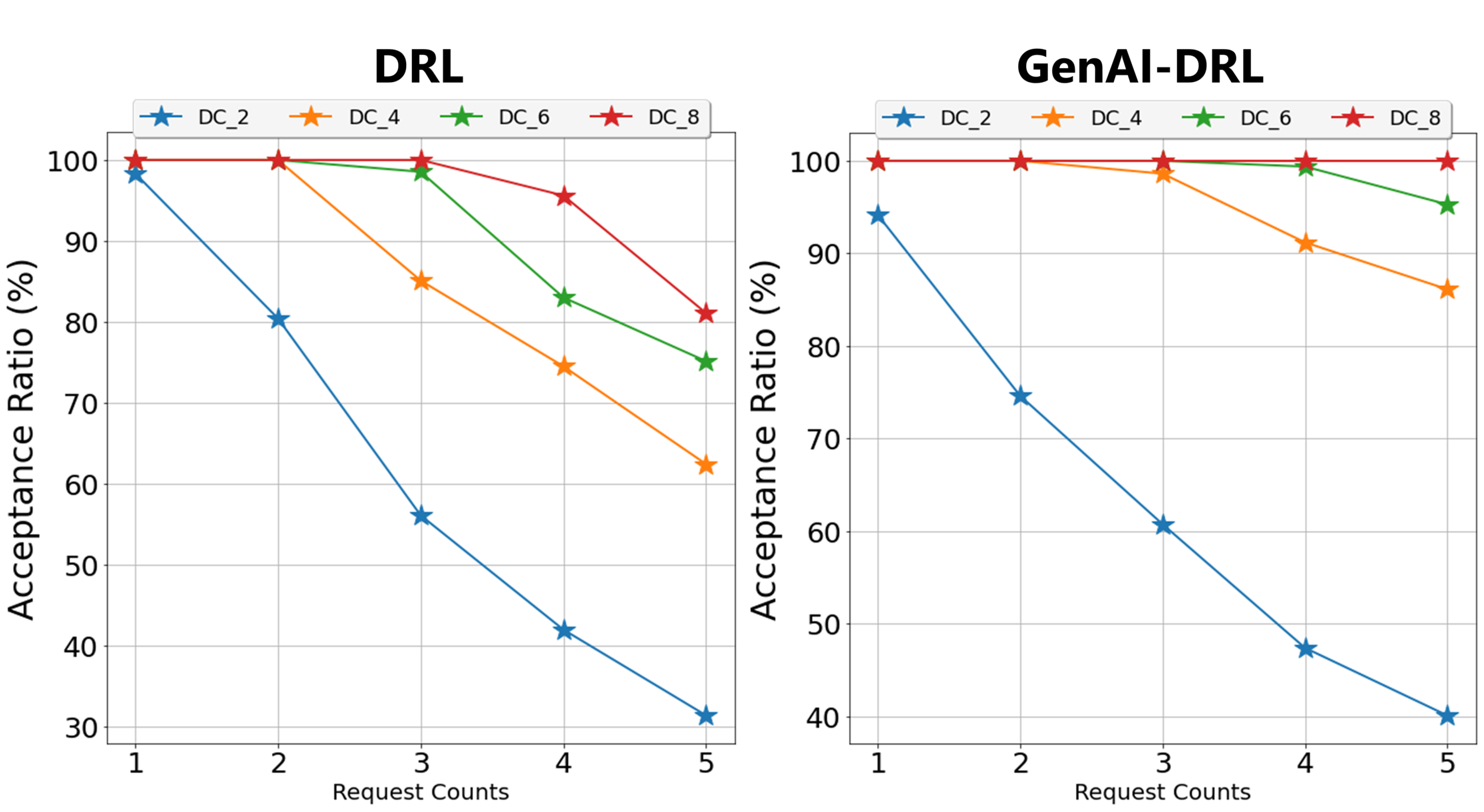}
    \caption{GenAI-DRL and DRL Comparison of SFC AccRatio Under Different DCs Number and Request Counts}
    \label{fig: res 2}
\end{figure}

\begin{figure}
    \centering
    \includegraphics[width=.90\linewidth]{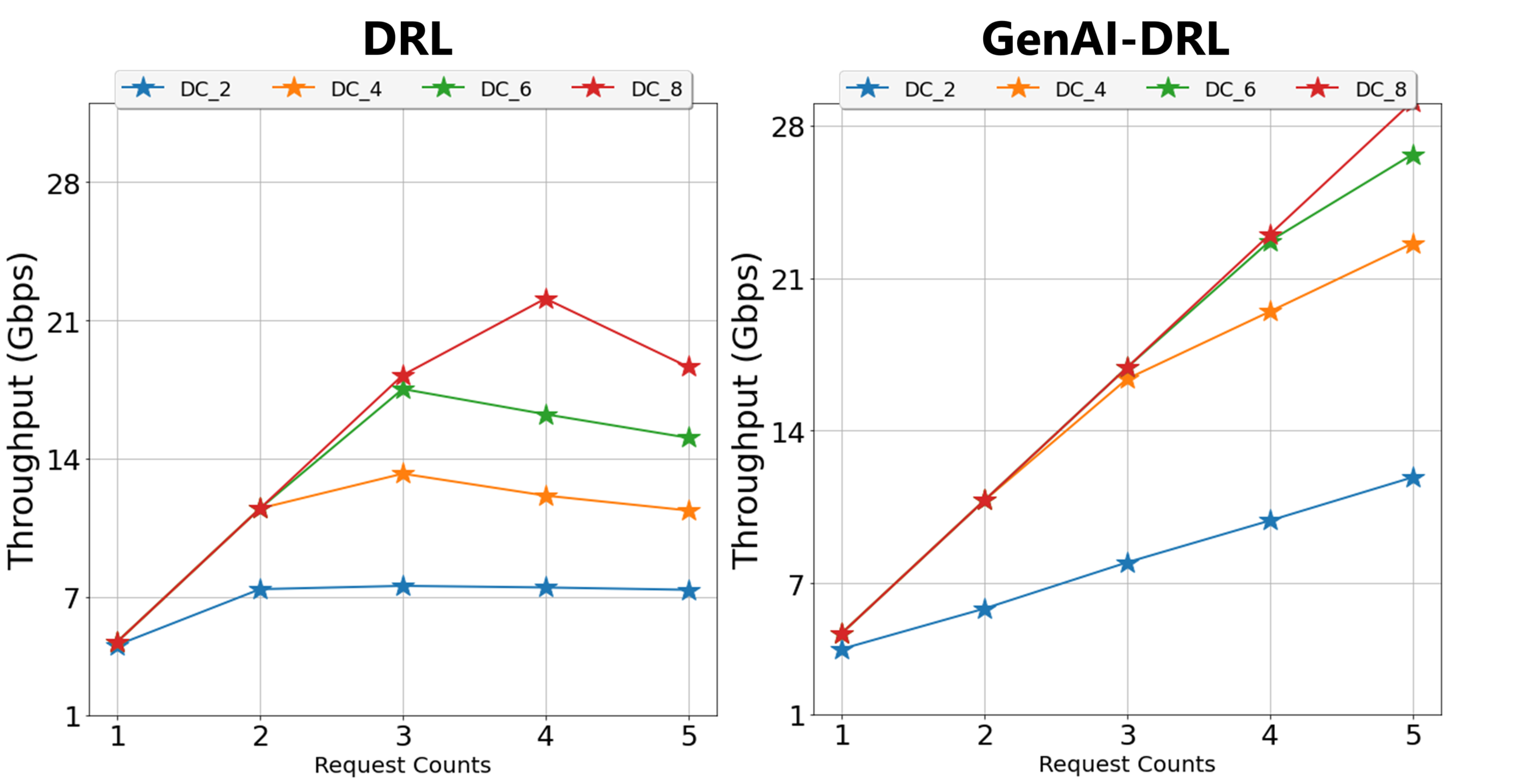}
    \caption{GenAI-DRL and DRL Comparison of Throughput Under Different DCs Number and Request Counts}
    \label{fig: res 3}
\end{figure}

SFC AccRatio and throughput comparison for DRL and GenAI-DRL models are given in Figs. \ref{fig: res 2} and \ref{fig: res 3}. Both figures show the performance comparison in various numbers of DCs, 2, 4, 6, and 8 and various request counts from 1 to 5. Note that each request count consists of all types of SFCs requests in bundles as per the size in Table \ref{tab:1}.

In \figurename \ref{fig: res 2}, the left part shows the SFC AccRatio of the DRL model, while the right part depicts the GenAI results. Though the DRL model can maintain its performance with maximum efficiency in 8 DCs till receiving 3 request counts, in other network configurations, there are some performance reductions once receiving more SFC requests. DRL model achieves 81\% of AccRatio in 5 SFC request counts with 8 DC while this score reduces to 76\%, 62\%, and 31.5\% in DCs 6, 4, and 2 environments, respectively. On the other hand, GenAI-DRL maintains its maximum performance for all request counts in 8 DC environments, and the other configurations, having 6, 4, and 2 DCs, achieve higher scores compared to the DRL model, which are 96\%, 87\%, and 40\% for request count 5. Though the proposed method's initial performance for DC 2 with slightly lower for request number 1 compared to DRL, it maintains higher performance in greater requests due to its planning ability over more massive demands.

In \figurename \ref{fig: res 3}, the left part shows the network throughput (Gbps) of the DRL model, while the right part illustrates the GenAI results. For DRL model,  throughput increases until request count 4 in 8 DCs case. In  6, 4, and 2 DCs cases,  since the AccRatio of Ind 4.0, AR, and MIoT, which require more bandwidth, are low for request counts 4 and 5, maximum throughput is achieved at request count 3.  On the other hand, the throughput of the proposed GenAI-DRL model keeps increasing in every configuration, and the maximum is achieved with more than 28 Gbps with 8 DC, while the minimum is around 10 Gbps for request number 5. Though SFC request rate reduction increases in higher request count as shown in \figurename \ref{fig: res 2}, the main reason it doesn't impact the throughput dramatically is that the proposed method effectively plans to satisfy the SFCs requests with high bandwidth requirements.

\section{Conclusions} \label{sec:6}

This research has explored the assistance of generative AI and its benefit to DRL in handling SFC requests under different network configurations with massive amounts of demands. A custom simulation with various DCs, connections, and SFC requests was created to collect data and train the generative model, and the trained model was used alongside the DRL model to satisfy SFC demands while the generative model handles DC selection for DRL model action. With the better planning and action selection, the proposed method GenAI-DRL outperforms DRL and benchmark heuristic models in terms of SFC AccRatio, E2E latency and throughput. The ongoing work focuses on the scalability and SFC provisioning under a larger network.

\section*{Acknowledgment }\label{Section6}
This work is supported in part by the Natural Sciences and Engineering Research Council of Canada (NSERC) Alliance Program, in part by the   MITACS Accelerate Program, and in part by the NSERC CREATE TRAVERSAL program.

%----------------------------------------------------------------------------------------
%	BIBLIOGRAPHY
%----------------------------------------------------------------------------------------

%\bibliography{biblio.bib}{}
\bibliographystyle{IEEEtran}

% \printbibliography[title={Bibliography}] % Print the

\end{document}